\documentclass[conference,a4paper]{IEEEtran}
\IEEEoverridecommandlockouts

\usepackage{cite}
\usepackage{amsmath,amssymb,amsfonts}
\usepackage{algorithm}
\usepackage{algorithmic}
\usepackage{graphicx}
\usepackage{textcomp}
\usepackage{xcolor}
\usepackage{multirow}
\usepackage{eso-pic}

\def\BibTeX{{\rm B\kern-.05em{\sc i\kern-.025em b}\kern-.08em
    T\kern-.1667em\lower.7ex\hbox{E}\kern-.125emX}}
\begin{document}
\bstctlcite{BSTcontrol}

\title{Joint Activation and Weight Quantization for Learned Image Compression}

\author{\IEEEauthorblockN{Yile Tu}
\IEEEauthorblockA{\textit{Yokohama National University, Japan} \\
tu-yile-cm@ynu.jp}
\and
\IEEEauthorblockN{Heming Sun}
\IEEEauthorblockA{\textit{Institute of Science Tokyo, Japan} \\
hemingsun@ieee.org}
\thanks{Corresponding author: Heming Sun.

This work was supported in part by JSPS KAKENHI Grant Number JP26K14754, and in part by JST, CRONOS, Japan Grant Number JPMJCS25N2.
}
}

\AddToShipoutPictureFG*{
  \AtPageUpperLeft{
    \put(\LenToUnit{0.5\paperwidth},\LenToUnit{-6mm}){
      \makebox[0pt][c]{
        \parbox[t]{\textwidth}{
          \centering\fontsize{7pt}{8pt}\selectfont
          \textcopyright~2026 IEEE. Personal use of this material is permitted. Permission from IEEE must be obtained for all other uses, in any current or future media, including reprinting/republishing this material for advertising or promotional purposes, creating new collective works, for resale or redistribution to servers or lists, or reuse of any copyrighted component of this work in other works.
        }
      }
    }
  }
}

\maketitle

\begin{abstract}
Learned image compression (LIC) models achieve competitive rate-distortion performance with increasingly large networks, making network quantization important for efficient deployment. Directly quantizing LIC models remains difficult in post-training quantization (PTQ) because compression performance is highly sensitive to quantization noise. Existing methods typically focus on optimizing either the weights or the activations, resulting in substantial rate-distortion degradation. To address this limitation, we introduce a two-stage pipeline that uses SmoothQuant to balance activation and weight ranges, followed by GPTQ to adapt the transformed weights. The proposed joint search framework optimizes the balancing transformation and adapted weights under the rate-distortion objective, accounting for the coupling between the two stages. Experiments across multiple LIC architectures and activation quantization configurations show that the joint search strategy consistently reduces rate-distortion degradation after quantization.
\end{abstract}


\section{Introduction}
Learned image compression (LIC) has achieved rate-distortion (RD) performance competitive with traditional image codecs through increasingly expressive neural network architectures \cite{balle2018variational, cheng2020learned, ho2021anfic, liu2023learned, li2024frequency}, and has been adopted into emerging JPEG AI standardization efforts \cite{alshina2024jpeg, esenlik2025overview}. At the same time, the growing complexity of modern LIC models increases memory overhead and computational cost during inference, limiting efficient deployment on practical hardware platforms. Network quantization reduces these costs by converting floating-point weights and activations into low-bit integer representations suitable for fixed-point inference on hardware accelerators \cite{sun2022f, sun2024fpga, chen2025exploring}. Quantization is usually performed through either quantization-aware training (QAT) or post-training quantization (PTQ). While QAT requires resource-intensive retraining, PTQ offers a highly efficient alternative by calibrating already trained models using a small representative dataset.

Standard PTQ often maintains classification accuracy with minor degradation in models such as ResNet, whereas the same quantization setting can cause severe coding-efficiency loss when applied to LIC. The different behavior arises because image compression is a low-level task that jointly optimizes reconstruction distortion and bitrate, making LIC more sensitive to quantization than standard classification models \cite{yang2025subset}. Quantization errors can also be amplified through the encoder-decoder transform and affect both reconstruction distortion and estimated bitrate.

Several PTQ methods have been developed for LIC and achieve promising rate-distortion performance \cite{Shi2022RateDistortionOP, sun2025q, yang2025subset, yu2025activation}. However, they usually focus on optimizing either the activation side or the weight side, which may leave the other side insufficiently addressed. To address this limitation, our method first balances the activation and weight distributions and then applies adaptive weight quantization. The balancing step shifts part of the quantization burden from activations to weights, while the adaptive quantization step further reduces weight-side error. In this paper, we instantiate this pipeline using SmoothQuant \cite{xiao2023smoothquant} and GPTQ \cite{Frantar2022GPTQP}. To better consider the coupling between the transformation and adaptation stages and optimize the rate-distortion objective, we propose a joint search framework. For each module, the framework evaluates candidate transformation parameters together with the corresponding adapted weights, selecting the candidate that minimizes the rate-distortion loss on calibration data. The main contributions of this work are summarized as follows:

\begin{itemize}
\item We introduce a two-stage PTQ pipeline for LIC that performs a balancing transformation via SmoothQuant followed by adaptive weight quantization via GPTQ.
\item We propose a joint search framework that optimizes activation balancing and weight adaptation under the rate-distortion objective to address their coupling.
\item We evaluate the framework under diverse quantization configurations, consistently reducing rate-distortion degradation across multiple LIC models and datasets.
\end{itemize}

\section{Related Work}

PTQ is widely studied to quantize neural networks at low cost. AdaRound \cite{Nagel2020UpOD} learns weight rounding decisions by minimizing reconstruction error on calibration data. SmoothQuant \cite{xiao2023smoothquant} applies output-preserving channel scaling to balance activation and weight ranges before quantization. GPTQ \cite{Frantar2022GPTQP} uses second-order information to compensate weight quantization error. QuIP \cite{chee2023quip} further shows that equivalence transformations can be combined with adaptive weight optimization to reduce quantization error.

For LIC, prior work has studied quantization from several directions. Ballé et al.~\cite{Ball2019IntegerNF} studied integer networks to support deterministic inference and entropy coding. Sun et al.~\cite{sun2020end, sun2021learned} explored fixed-point quantization and fine-tuning strategies for such models. Under the post-training setting, RDO-PTQ~\cite{Shi2022RateDistortionOP} extends adaptive rounding to LIC by optimizing quantization parameters with respect to rate-distortion loss, but requires iterative fine-tuning on calibration data. Q-LIC~\cite{sun2025q} mitigates quantization error by dynamically splitting sensitive channels to reduce their dynamic range and pruning low-energy channels according to the rate-distortion objective. SSW~\cite{yang2025subset} introduces a non-uniform weight quantization scheme by selecting layer-specific subsets from a universal point set. MP-PTQ~\cite{yu2025mixed} studies activation quantization mainly under a mixed-precision setting. AWDB-PTQ~\cite{yu2025activation} migrates activation quantization difficulty to the weight side before quantization and reports fixed-precision W8A8 results, while also extending the framework to mixed-precision search.

\section{Methodology}
\subsection{Balancing and Adaptation Pipeline}
\subsubsection{Quantization Error Analysis}

To preserve model performance after quantization, a common objective is to minimize the local reconstruction error of individual layers as a proxy for the global task loss \cite{Nagel2020UpOD}. Consider a linear layer or unfolded convolution with weight matrix $\mathbf{W} \in \mathbb{R}^{C_{\mathrm{out}} \times C_{\mathrm{in}}}$ and input activation matrix $\mathbf{X} \in \mathbb{R}^{C_{\mathrm{in}} \times N}$. Let $\mathbf{\hat{W}}$ and $\mathbf{\hat{X}}$ denote their quantized counterparts, and define the quantization noises as $\Delta \mathbf{W} = \mathbf{\hat{W}} - \mathbf{W}$ and $\Delta \mathbf{X} = \mathbf{\hat{X}} - \mathbf{X}$. Assuming weight and activation quantization noises are independent and zero-mean, and neglecting the higher-order joint error term $\Delta \mathbf{W}\Delta \mathbf{X}$, the local reconstruction MSE can be approximated as:
\begin{align}
\mathcal{L}_{\mathrm{local}}
&= \mathbb{E} \left[ \| \mathbf{\hat{W}}\mathbf{\hat{X}} - \mathbf{W}\mathbf{X} \|_F^2 \right] \nonumber \\
&\approx \mathbb{E} \left[ \| \Delta \mathbf{W} \mathbf{X} \|_F^2 \right]
+ \mathbb{E} \left[ \| \mathbf{W} \Delta \mathbf{X} \|_F^2 \right].
\end{align}

In several common LIC architectures, activations can exhibit greater sensitivity to quantization compared to weights \cite{yu2025activation}. Under these conditions, purely weight-adaptive PTQ methods are limited, as they optimize only the weight-induced error term while leaving the dominant activation-induced term largely unaddressed.

To balance these error components, an output-preserving balancing transformation matrix $\mathbf{T} \in \mathbb{R}^{C_{\mathrm{in}} \times C_{\mathrm{in}}}$ is introduced following the joint quantization paradigm \cite{liu2025comprehensive}. The transformed weights and activations are defined as
\begin{equation}
\mathbf{W}' = \mathbf{W}\mathbf{T}, \qquad
\mathbf{X}' = \mathbf{T}^{-1}\mathbf{X},
\end{equation}
which preserves the full-precision layer output, since $\mathbf{W}'\mathbf{X}'=\mathbf{W}\mathbf{X}$. Rather than eliminating quantization error, the balancing transformation redistributes quantization difficulty between activations and weights. The inverse transformation $\mathbf{T}^{-1}$ suppresses the activation range, while $\mathbf{T}$ rescales the weights.

To account for the subsequent adaptive weight quantization, let $\mathcal{A}(\cdot)$ denote an adaptive weight quantizer that quantizes the transformed weights using the transformed calibration activations. We define the adaptively quantized transformed weights as
$\hat{\mathbf{W}}'_{\mathcal{A}}
= \mathcal{A}(\mathbf{W}', \mathbf{X}')$, and the corresponding adaptive weight
quantization error as $\Delta \mathbf{W}'_{\mathcal{A}}
= \hat{\mathbf{W}}'_{\mathcal{A}} - \mathbf{W}'$.
Let $\Delta\mathbf{X}'=\hat{\mathbf{X}}'-\mathbf{X}'$ denote the transformed activation quantization noise.
Under the balancing transformation and adaptive weight quantization, the local reconstruction MSE can be approximated as
\begin{equation}
\mathcal{L}_{\mathrm{local}}(\mathbf{T})
\approx
\mathbb{E} \left[ \| \Delta\mathbf{W}'_{\mathcal{A}} \mathbf{X}' \|_F^2 \right]
+
\mathbb{E} \left[ \| \mathbf{W}' \Delta\mathbf{X}' \|_F^2 \right].
\end{equation}

\begin{figure}[t]
\centering
\includegraphics[width=\columnwidth]{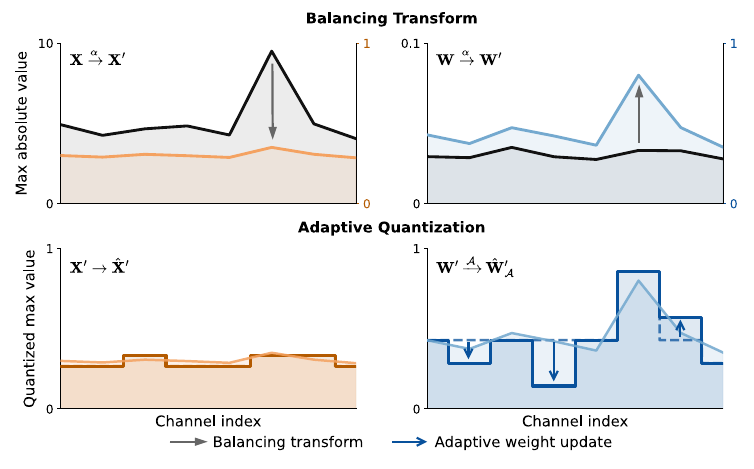}
\caption{Illustration of balancing activation and weight distributions before adaptive weight quantization. The balancing transformation suppresses activation range while shifting the corresponding scale to the weights.}
\label{fig:overview}
\end{figure}
\subsubsection{Instantiation with SmoothQuant and GPTQ}

To instantiate the two-stage pipeline, we use SmoothQuant \cite{xiao2023smoothquant} as the balancing transformer and GPTQ \cite{Frantar2022GPTQP} as the adaptive weight quantizer. Specifically, we restrict the transformation matrix to a diagonal scaling matrix $\mathbf{T}=\operatorname{diag}(\mathbf{s})$, where $\mathbf{s} \in \mathbb{R}^{C_{\mathrm{in}}}$ contains the channel-wise scaling factors. For each input channel $j$, the scaling coefficient $s_j$ is parameterized by a migration parameter $\alpha \in [0,1]$:
\begin{equation}
s_j = \max(|\mathbf{X}_{j, :}|)^{\alpha} / \max(|\mathbf{W}_{:, j}|)^{1-\alpha},
\end{equation}
where $\max(|\mathbf{X}_{j, :}|)$ and $\max(|\mathbf{W}_{:, j}|)$ denote the maximum absolute values of the activation channel and the corresponding weight column, respectively. The parameter $\alpha$ controls the quantization difficulty transferred from activations to weights, with a common SmoothQuant setting given by $\alpha=0.5$.

After applying the diagonal transformation, we instantiate the adaptive weight quantizer $\mathcal{A}$ with GPTQ. GPTQ quantizes the transformed weight matrix $\mathbf{W}'$ using the transformed calibration features $\mathbf{X}'$, and performs second-order error compensation by updating the remaining unquantized weights after each quantization step. Because both $\mathbf{W}'$ and $\mathbf{X}'$ are determined by the scaling parameter $\alpha$, the final quantized weights produced by GPTQ are also functions of $\alpha$, denoted as $\hat{\mathbf{W}}'_{\mathcal{A}}(\alpha)$. Consequently, changing $\alpha$ alters not only the activation and weight ranges induced by SmoothQuant, but also the subsequent adaptive weight updates.

\subsection{Joint Search Framework}
\subsubsection{Scaling Search and Objective Mismatch}

\begin{figure}[t]
\centering
\includegraphics[width=\columnwidth]{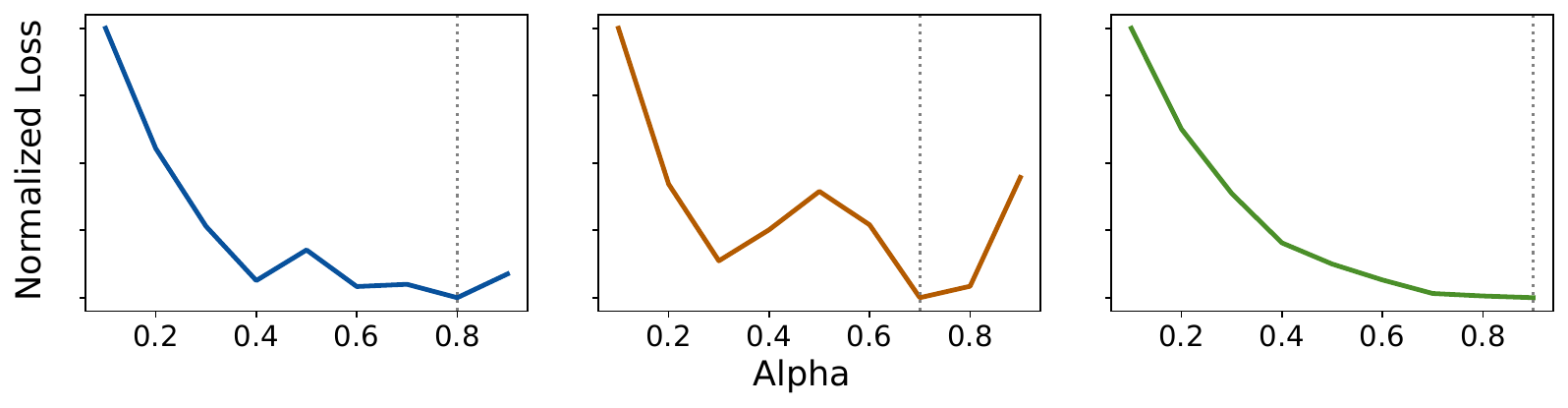}
\caption{Rate-distortion loss versus migration parameter $\alpha$ for representative module-wise search units under tensor-wise activation quantization. Each curve is min–max normalized.}
\label{fig:alpha_search}
\end{figure}

SmoothQuant commonly adopts $\alpha=0.5$, which evenly balances quantization difficulty between activations and weights under an idealized range-balancing view, but this setting is insufficient under practical hardware quantization settings. The SmoothQuant scaling is applied per input channel, whereas weights are quantized per output channel. The structural mismatch makes the optimal scaling search equivalent to a makespan scheduling problem, which remains NP-hard \cite{yu2025activation}. Adaptive weight quantization departs from the independent uniform-noise model by coupling weight error with the transformed calibration activations. Additionally, activation quantization error is strongly affected by non-linear noise propagated from upstream layers, which cannot be captured by standard static noise assumptions. As shown in Fig.~\ref{fig:alpha_search}, the minimizing migration parameter varies across representative search units.

Since the weights adapted by GPTQ also depend on $\alpha$, selecting $\alpha$ only from the SmoothQuant transformation stage may be mismatched with the final compression objective. The rate-distortion objective provides a global evaluation that includes the influence of each candidate on subsequent modules and directly reflects the target optimized by the compression model, while remaining practical for calibration. We therefore jointly optimize $\alpha$ and the adapted weights using the rate-distortion loss:
\begin{equation}
\mathcal{L}_{\mathrm{RD}} = D(\mathbf{x}, \hat{\mathbf{x}}) + \lambda R(\hat{\mathbf{y}})
\end{equation}
where $D$ represents the reconstruction distortion, $R$ is the estimated bitrate of the quantized latents $\hat{\mathbf{y}}$, and $\lambda$ is the trade-off parameter.

\subsubsection{Joint Search Implementation Details}

Instead of optimizing layers in isolation or using a single migration parameter for the entire network, we perform the joint search in a module-wise manner. We divide the network into five search units: the analysis transform $g_a$, synthesis transform $g_s$, hyper-analysis transform $h_a$, hyper-synthesis transform $h_s$, and the remaining entropy-related modules. Module-wise calibration avoids overly greedy layer-wise decisions while providing finer control than a single global parameter and remaining substantially more efficient than layer-wise calibration.

For each search unit $U_k$, we assign a single migration parameter $\alpha_k$ shared by all transformed layers inside that unit, and select it from a predefined uniform grid $\mathcal{G}_\alpha=\{0.1,0.2,\dots,0.9\}$. We adopt this simple grid search to show that the performance gain is attributable to the proposed joint search framework rather than the search strategy. During each candidate evaluation for $U_k$, previously calibrated units retain their committed quantization configurations. The rate-distortion evaluation therefore accounts for upstream quantization noise accumulated from previously calibrated units.

Within the joint search framework, percentile clipping is used to reduce the influence of extreme activation outliers during activation-range calibration \cite{li2019fully}. We set the clipping threshold to the 99.999-th percentile in our implementation. To integrate GPTQ with the SmoothQuant scaling, we first apply damping regularization to the raw activation Hessian, and then incorporate the SmoothQuant scaling into the inverse-Hessian factor by multiplying it with $\operatorname{diag}(\mathbf{s})$. This keeps the damping term tied to the original calibration activations while providing the scaled inverse-Hessian factor used by GPTQ under the transformed activations. The complete procedure of the joint search framework is summarized in Algorithm~\ref{alg:joint_search}.

\begin{table}[t]
\centering
\refstepcounter{algorithm}
\label{alg:joint_search}

\newcounter{jointsearchline}
\setcounter{jointsearchline}{0}
\newcommand{\jointsearchlinenum}{\stepcounter{jointsearchline}\thejointsearchline:}
\centering
\fontsize{8.2pt}{10pt}\selectfont
\renewcommand{\arraystretch}{1.1}
\begin{tabular}{l p{\dimexpr \columnwidth - 3em - 2\tabcolsep \relax}}
\hline\hline
\multicolumn{2}{l}{\textbf{Algorithm~\thealgorithm: Proposed Joint Search Framework}} \\
\hline
\multicolumn{2}{p{\dimexpr \columnwidth - 2\tabcolsep \relax}}{
\textbf{Input:} Pre-trained model $M$, calibration set $\mathcal{X}$, search grid $\mathcal{G}_\alpha$
} \\
\multicolumn{2}{p{\dimexpr \columnwidth - 2\tabcolsep \relax}}{
\textbf{Output:} Calibrated quantized model $M_q$
} \\
\hline
\jointsearchlinenum & Group layers into module-wise search units $\mathcal{U}$ \\
\jointsearchlinenum & \textbf{for} each module-wise search unit $U_k \in \mathcal{U}$ \textbf{do} \\
\jointsearchlinenum & \quad \textit{// Observation Phase} \\
\jointsearchlinenum & \quad Collect per-channel ranges
$\max(|\mathbf{X}_{j,:}|)$ and $\max(|\mathbf{W}_{:,j}|)$ \\
\jointsearchlinenum & \quad Collect Hessian
$\mathbf{H} = \mathbf{X}\mathbf{X}^\top$ \\
\jointsearchlinenum & \quad Apply GPTQ damping
$\mathbf{H}_d \leftarrow
\mathbf{H}
+ \gamma\,\mathrm{mean}(\mathrm{diag}(\mathbf{H}))\mathbf{I}$ \\
\jointsearchlinenum & \quad Compute inverse Cholesky factor $\mathbf{L}$ s.t.
$\mathbf{L}^\top \mathbf{L} \approx \mathbf{H}_d^{-1}$ \\
\jointsearchlinenum & \quad \textit{// Trial Search Phase} \\
\jointsearchlinenum & \quad \textbf{for} each candidate
$\alpha \in \mathcal{G}_\alpha$ \textbf{do} \\
\jointsearchlinenum & \quad\quad
$s_j \leftarrow
\max(|\mathbf{X}_{j,:}|)^\alpha /
\max(|\mathbf{W}_{:,j}|)^{1-\alpha}$ \\
\jointsearchlinenum & \quad\quad
$\mathbf{X}' \leftarrow
\mathrm{diag}(\mathbf{s})^{-1}\mathbf{X}$,
$\mathbf{W}' \leftarrow
\mathbf{W}\mathrm{diag}(\mathbf{s})$
\quad \textit{// SmoothQuant} \\
\jointsearchlinenum & \quad\quad
$\mathbf{L}' \leftarrow
\mathbf{L}\mathrm{diag}(\mathbf{s})$
\quad \textit{// inverse-Hessian update} \\
\jointsearchlinenum & \quad\quad
$\hat{\mathbf{W}}'_{\mathcal{A}}
\leftarrow
\text{ApplyGPTQ}(U_k,\mathbf{W}',\mathbf{L}')$
\textit{// adaptive quantization} \\
\jointsearchlinenum & \quad\quad
$\mathcal{L}_{\mathrm{RD}}(\alpha)
\leftarrow
\text{EvaluateRD}
(M,\alpha,\hat{\mathbf{W}}'_{\mathcal{A}})$
on $\mathcal{X}$ \\
\jointsearchlinenum & \quad\quad
Reset $U_k$ to floating-point state for next trial \\
\jointsearchlinenum & \quad \textbf{end for} \\
\jointsearchlinenum & \quad
$\alpha_k^*
\leftarrow
\arg\min_{\alpha \in \mathcal{G}_\alpha}
\mathcal{L}_{\mathrm{RD}}(\alpha)$ \\
\jointsearchlinenum & \quad \textit{// Commit Phase} \\
\jointsearchlinenum & \quad
Apply optimal
$(\alpha_k^*,\hat{\mathbf{W}}'_{\mathcal{A}}{}^{*})$
to the search unit $U_k$ \\

\jointsearchlinenum & \textbf{end for} \\

\jointsearchlinenum & \textbf{return} Calibrated quantized model $M_q$ \\

\hline\hline

\end{tabular}
\end{table}

\section{Experiments}

\subsection{Experimental Setup}

We evaluate our joint search framework on representative learned image compression models. We quantize the weights and inputs of all convolutional and linear layers, as well as the inputs of matrix multiplications in transformer-based blocks. Weights are quantized using channel-wise symmetric uniform quantization, while activations are quantized using either tensor-wise or channel-wise symmetric uniform quantization. The input to the first layer is quantized using unsigned quantization because its values are nonnegative. Percentile clipping is applied for W8A8 and W9A9, and omitted for W10A10 due to the sufficient resolution of the larger bit-width. 

We use official pre-trained checkpoints from public repositories. Results for prior LIC quantization methods are quoted from their original publications. To prevent overfitting, calibration is performed using 32 ImageNet~\cite{deng2009imagenet} images randomly cropped to $256 \times 256$ pixels. Experiments used an NVIDIA RTX PRO 6000 GPU. We measure rate-distortion performance by BD-rate increase over the floating-point baseline on Kodak~\cite{kodak}, Tecnick~\cite{tecnick}, and CLIC~\cite{CLIC2020}, depending on the baselines compared.

\begin{figure}[t]
\centering
\includegraphics[width=0.9\columnwidth]{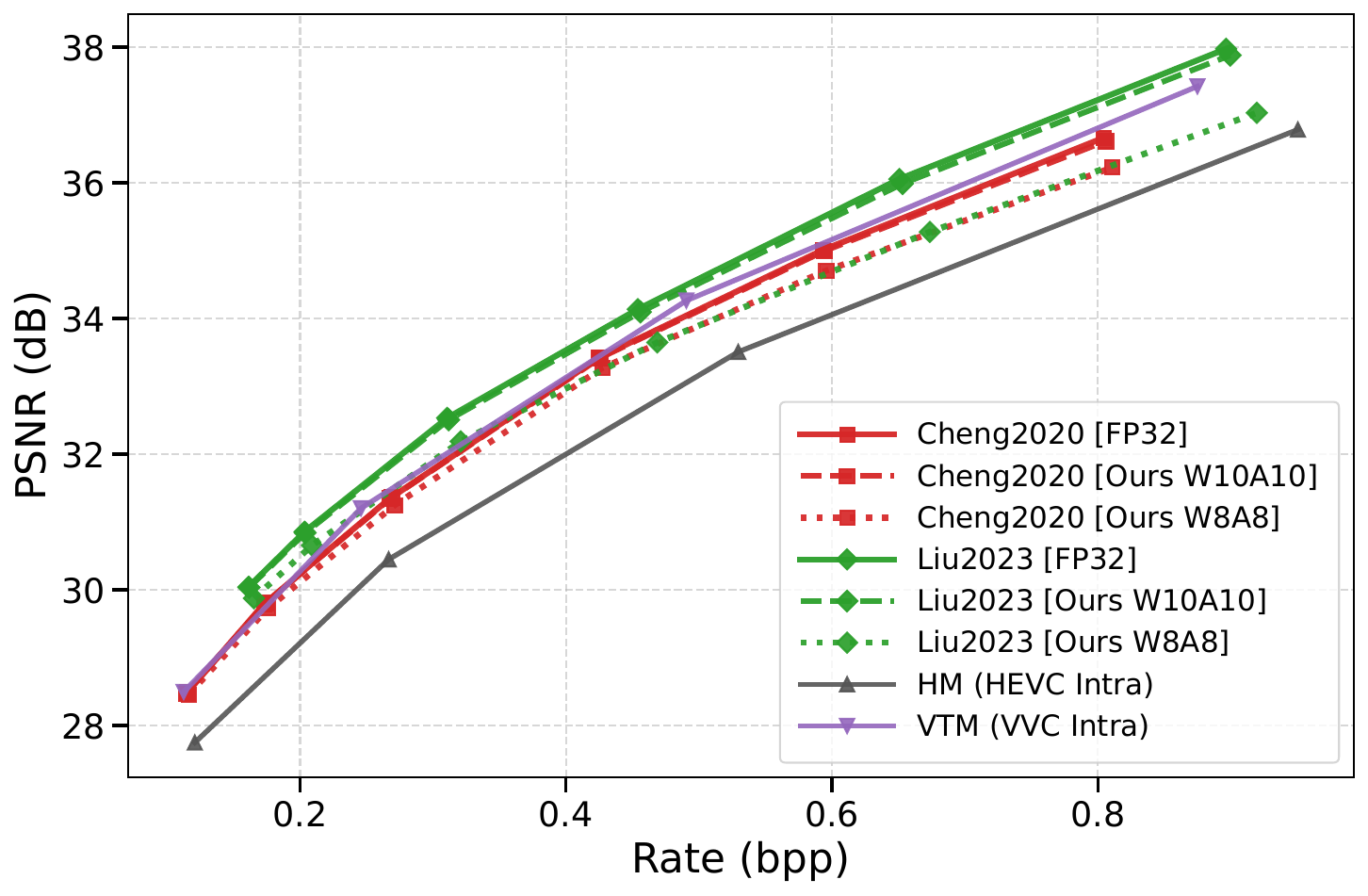}
\caption{Rate-distortion curves on Kodak for Cheng2020 and Liu2023 under tensor-wise activation quantization, with HM and VTM anchor curves obtained from CompressAI~\cite{begaint2020compressai}.}
\label{fig:rd_curves}
\end{figure}

\begin{table}[t]
\centering
\caption{BD-rate increase (\%) over floating-point models under tensor-wise activation quantization.}
\label{tab:tensor_multibit}
\renewcommand{\arraystretch}{1.1}
\setlength{\tabcolsep}{1.75pt}

\begin{tabular}{c l c c c c c c}
\hline\hline
\multirow{2}{*}{\textbf{Bit-width}} &
\multirow{2}{*}{\textbf{Method}} &
\multicolumn{2}{c}{\textbf{Ball\'e2018}} &
\multicolumn{2}{c}{\textbf{Cheng2020}} &
\multicolumn{2}{c}{\textbf{Liu2023}} \\
\cline{3-8}
& & \textbf{Kodak} & \textbf{CLIC}
& \textbf{Kodak} & \textbf{CLIC}
& \textbf{Kodak} & \textbf{CLIC} \\
\hline

\multirow{3}{*}{W8A8} 
& SmoothQuant \cite{xiao2023smoothquant}
& 155.56 & 259.07
& 12.01 & 20.37
& 34.66 & 55.32 \\
& AWDB-PTQ \cite{yu2025activation}
& 61.83 & 105.92
& 8.74 & 13.06
& 21.60 & 31.93 \\
& \textbf{Ours}
& \textbf{21.15} & \textbf{31.52}
& \textbf{4.68} & \textbf{8.33}
& \textbf{14.69} & \textbf{21.77} \\

\hline

\multirow{3}{*}{W9A9} 
& SmoothQuant \cite{xiao2023smoothquant}
& 41.19 & 66.80
& 2.98 & 5.56
& 7.88 & 11.39 \\
& AWDB-PTQ \cite{yu2025activation}
& 16.92 & 27.04
& 2.85 & 3.77
& 6.79 & 7.31 \\
& \textbf{Ours}
& \textbf{6.84} & \textbf{9.52}
& \textbf{1.39} & \textbf{3.11}
& \textbf{3.95} & \textbf{6.28} \\

\hline

\multirow{3}{*}{W10A10} 
& SmoothQuant \cite{xiao2023smoothquant}
& 10.63 & 16.05
& 0.64 & 1.40
& 1.95 & 2.90 \\
& AWDB-PTQ \cite{yu2025activation}
& 5.19 & 8.02
& 1.42 & 1.26
& 3.46 & 2.54 \\
& \textbf{Ours}
& \textbf{3.79} & \textbf{5.81}
& \textbf{0.33} & \textbf{1.17}
& \textbf{1.24} & \textbf{2.04} \\

\hline\hline
\end{tabular}
\end{table}

\begin{table}[t]
\centering
\caption{BD-rate increase (\%) for Cheng2020 under W8A8 channel-wise activation quantization. (`--' denotes unreported results).}
\label{tab:channel_metrics}
\renewcommand{\arraystretch}{1.12}
\setlength{\tabcolsep}{4.8pt}
\begin{tabular}{l c c c c}
\hline\hline
\multirow{2}{*}{\textbf{Method}}
& \multicolumn{2}{c}{\textbf{MSE}}
& \multicolumn{2}{c}{\textbf{MS-SSIM}} \\
\cline{2-5}
& \textbf{Kodak} & \textbf{Tecnick}
& \textbf{Kodak} & \textbf{Tecnick} \\
\hline

RDO-PTQ \cite{Shi2022RateDistortionOP}
& 4.88 & 6.86 & 4.65 & 5.22 \\

Q-LIC \cite{sun2025q}
& 10.50 & -- & 4.40 & -- \\

SSW \cite{yang2025subset}
& 4.05 & -- & -- & -- \\

\textbf{Ours}
& \textbf{3.60} & \textbf{6.59}
& \textbf{2.94} & \textbf{3.75} \\

\hline\hline
\end{tabular}
\end{table}

\subsection{Comparison with Prior PTQ Methods}

We evaluate our method under tensor-wise activation quantization, as this granularity represents the standard for commodity hardware accelerators due to its high computational efficiency \cite{Nagel2021AWP}. Under this setting, we compare our framework against SmoothQuant and AWDB-PTQ using Ball{\'e}2018~\cite{balle2018variational}, Cheng2020~\cite{cheng2020learned}, and Liu2023~\cite{liu2023learned}. The BD-rate results on the Kodak and CLIC datasets across multiple bit-widths are summarized in Table~\ref{tab:tensor_multibit}, and the rate-distortion curves are visualized in Fig.~\ref{fig:rd_curves}. The joint search framework consistently reduces rate-distortion degradation compared to the baseline methods across W8A8, W9A9, and W10A10 settings.

We also conduct experiments under channel-wise activation quantization, which is occasionally implemented in specialized hardware such as FPGAs to preserve precision. Our unified pipeline supports both coarse and fine-grained activation structures. Under this setting, we evaluate Cheng2020~\cite{cheng2020learned} under W8A8 quantization for models optimized with both MSE and MS-SSIM objectives. As shown in Table~\ref{tab:channel_metrics}, the proposed joint search framework yields lower rate-distortion overhead than previous methods under channel-wise activation quantization. The channel-wise advantage shows that the framework remains effective under fine-grained activation quantization.

\begin{table}[t]
\centering
\caption{BD-rate increase (\%) over floating-point models on Kodak under W8A8 quantization with different activation quantization granularities.}
\label{tab:ablation_kodak}
\renewcommand{\arraystretch}{1.12}
\setlength{\tabcolsep}{4pt}
\begin{tabular}{l cc cc}
\hline\hline
\textbf{Variant}
& \multicolumn{2}{c}{\textbf{Tensor-wise}}
& \multicolumn{2}{c}{\textbf{Channel-wise}} \\
\cline{2-5}
& \textbf{Cheng2020} & \textbf{Liu2023}
& \textbf{Cheng2020} & \textbf{Liu2023} \\
\hline
RTN          & 17.43 & 79.00 & 6.24 & 16.61 \\
SmoothQuant \cite{xiao2023smoothquant} & 7.87  & 21.92 & 4.69 & 12.88 \\
GPTQ \cite{Frantar2022GPTQP}         & 15.08 & 70.46 & 4.05 & 11.04 \\
SmoothQuant+GPTQ      & 7.08  & 19.46 & 3.88 & 10.57 \\
\hline
Ours w/ MSE Search   & 4.71  & 14.77 & 3.81 & 10.47 \\
Ours w/o Percentile & 5.46  & 20.42 & 3.92 & 11.53 \\
\textbf{Ours} & \textbf{4.68} & \textbf{14.69} & \textbf{3.60} & \textbf{10.44} \\
\hline\hline
\end{tabular}
\end{table}

\subsection{Ablation Studies}

In Table~\ref{tab:ablation_kodak}, MSE Search selects the migration parameter using local reconstruction MSE rather than rate-distortion loss, while w/o Percentile disables percentile clipping for activation-range calibration. All component baselines use the same percentile-clipped activation ranges as our method. GPTQ alone remains close to round-to-nearest (RTN) under tensor-wise quantization but becomes much more effective under channel-wise quantization, showing that the allocation of quantization difficulty differs across activation granularities. The proposed framework remains robust under these different activation and weight sensitivity patterns. The gap between SmoothQuant+GPTQ and our method is especially clear under tensor-wise activation quantization, demonstrating the benefit of jointly selecting the balancing and adaptation configuration for improved rate-distortion recovery.

\section{Conclusion}
This paper introduced a joint search framework for PTQ of LIC models. The framework combines a balancing transformation with adaptive weight quantization and optimizes both under the compression objective. Experiments under diverse quantization configurations and LIC architectures demonstrate that the proposed method effectively reduces rate-distortion degradation caused by quantization. Future work will explore quantization-aware training with balancing transformations between activation and weight distributions to further improve learned image compression.

\bibliographystyle{IEEEtran}
\bibliography{references}

\end{document}